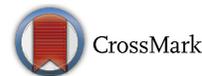

REGULAR PAPER

# A local hidden-variable model for experimental tests of the GHZ puzzle

Brian R. La Cour

**Abstract** The Greenberger–Horne–Zeilinger (GHZ) puzzle has been used to study quantum nonlocality and provide an all-or-nothing, no-go theorem for local hidden-variable models. Recent experiments using coincident-detected entangled photons prepared in a three-particle GHZ state have been used to test quantum nonlocality, but fail to rule out local realism due to a reliance on the fair-sampling hypothesis and insufficient detection efficiency. This paper describes a physically motivated local hidden-variable model based on amplitude-threshold detection that is capable of producing similar results. Detection efficiencies for the model are within the bounds permitted for local realism and, interestingly, exhibit statistical correlations between detectors, even when the detection events are spacelike separated. Increasing the detection threshold improves agreement with the ideal quantum predictions at the cost of decreased detection efficiency. A curious emergent feature of the model is that detection efficiencies may depend upon which observables are chosen for measurement.

## 1 Introduction

The concept of quantum nonlocality has fascinated physicists ever since the seminal paper of Einstein et al. [1]. Inspired by this work, Bell derived his now-famous inequality regarding correlations between pairs of observables, one that must be satisfied by any local hidden-variable model [2]. Numerous tests of variants of Bell's inequality have been performed experimentally by now, and almost all of them observe a violation, as predicted by quantum mechanics [3–12].

Common to many optical tests of Bell's inequality is the so-called "detection loophole", which arises from the necessity to rely on coincident photon detection events, as undetected photons cannot be measured directly. Assuming that detected photons are representative of all incident photons amounts to adopting the fair-sampling hypothesis, which is used implicitly when one normalizes by the total number of coincident-detected events. This procedure may, however, allow for a local hidden-variable interpretation, as first demonstrated by Pearle [13]. In particular, for detection efficiencies at or below a certain critical value, local hidden-variable models can reproduce quantum predictions exactly [14,15].

B. R. La Cour (✉)
Applied Research Laboratories, The University of Texas at Austin, P.O. Box 8029, Austin, TX 78713-8029, USA
e-mail: blacour@arlut.utexas.edu

A particular variant of Bell's argument was introduced by Greenberger, Horne, and Zeilinger using what is now known as a three-particle GHZ state [16,17]. This result was later recast by Mermin as an inequality to be satisfied by any local hidden-variable model [18]. Mermin's derivation assumed that all measurements result in a detection, but a local hidden-variable model with imperfect detection efficiency may still violate the inequality. Larsson has shown that, under quite general conditions, a triple-coincidence detection efficiency above 75 % is needed to preclude the existence of a local hidden-variable model capable of reproducing the perfect correlations predicted for a three-particle GHZ state [19]. This result was later generalized by Cabello et al. for $n$-particle GHZ states, albeit with the assumption of independent missed-detection events [20]. Thus, when detection efficiencies are sufficiently low a local hidden-variable interpretation may still be maintained.

The first experimental test of the GHZ puzzle was performed by Pan et al. [21] using triple-coincidence-detected entangled photons produced through parametric down conversion. Their results were in good agreement with quantum predictions, but low detection efficiencies and use of the fair-sampling hypothesis left open the detection loophole. More recently, Erven et al. have performed an experiment showing a violation of the Mermin inequality using independent, spacelike separated measurements, although with efficiencies below the Larsson bound [22]. In a related experiment, Hamel et al. have performed an experiment showing violations of both the Mermin and Svetlichny inequalities using high-efficiency superconducting photodetectors but with a ratio of triple to double coincidence rates that, again, was well below the 75 % Larsson bound [23]. We may conclude that none of these experiments have succeeded in ruling out a local hidden-variable interpretation of the experimental observations.

It is, however, difficult to accept a hidden-variable interpretation of these experiments without an intuitive understanding of how the fair-sampling hypothesis might reasonably be violated. One can easily construct various mathematical contrivances that serve to prove the existence of a local hidden-variable model, but these lack the generality and physical plausibility needed to be taken at all seriously. Furthermore, they typically provide no insight into the underlying physics, nor do they provide new observational predictions.

The present paper attempts to address these concerns by providing an intuitive understanding of how the fair-sampling hypothesis might reasonably be violated in such experiments through the construction and analysis of a specific local hidden-variable model. Specifically, we show that by considering only three-way coincident events, as was done in the aforementioned experiments, one can achieve a violation of the Mermin inequality while maintaining local realism. Missed-detection events are an implicit feature of the model, which illustrates how even ideal detectors can fail to register a detection. An interesting emergent feature of the model is that missed-detection events can be correlated even under strict locality conditions.

It is further shown that violations of the Mermin inequality may be made arbitrarily large, up to the maximum GHZ-state violation, by adjusting one of the model parameters (namely, the detection threshold) but only at the cost of decreased detection efficiency. These results are consistent with those of both Larsson and Cabello et al. since the efficiencies are within the bounds prescribed for local hidden-variable models. The model described here is based on previous work concerning bipartite states [24,25]; here it is applied and extended to local measurements on tripartite entangled states.

The organization of the paper is as follows: Sect. 2 provides a review of the quantum GHZ puzzle and discusses the role of non-detection events in its understanding. Section 3 describes a detection-based local hidden-variable model and compares its predictions to recent experimental observations. Section 4 examines how varying the detection threshold affects agreement with the quantum predictions and impacts detection efficiency. Conclusions are summarized in Sect. 5.

## 2 The GHZ puzzle

We begin with a description of the GHZ state in terms of the polarization modes of three photons. We use $|H\rangle$ and $|V\rangle$ to describe the $H/V$ basis of horizontal and vertical polarization modes, respectively, of a single photon and consider the three-photon GHZ state given by

$$|\psi\rangle = \frac{1}{\sqrt{2}}\Big[|HHH\rangle - i\,|VVV\rangle\Big]. \tag{1}$$

Upon preparation of this state, measurements are performed by each of three agents, Alice, Bob, and Charlie, who measure the left, middle, and right photons, respectively. Each agent selects one of two bases in which to measure: either the *R/L* basis, given by $|R\rangle = \frac{1}{\sqrt{2}}[|H\rangle + i\,|L\rangle]$ and $|L\rangle = \frac{1}{\sqrt{2}}[|H\rangle - i\,|L\rangle]$, or the *D/A* basis, given by $|D\rangle = \frac{1}{\sqrt{2}}[|H\rangle + |L\rangle]$ and $|A\rangle = \frac{1}{\sqrt{2}}[|H\rangle - |L\rangle]$. An outcome of $|R\rangle$ or $|D\rangle$ is reported as $+1$, while an outcome of $|L\rangle$ or $|A\rangle$ is reported as $-1$. In the absence of missed detections, no other outcomes are taken to be possible.

For these six measurements, we may define the following projection operators:

$$\mathbf{P}_A = \sum_{yz} |Ryz\rangle\langle Ryz|, \quad \mathbf{P}'_A = \sum_{yz} |Dyz\rangle\langle Dyz|, \tag{2a}$$

$$\mathbf{P}_B = \sum_{xz} |xRz\rangle\langle xRz|, \quad \mathbf{P}'_B = \sum_{xz} |xDz\rangle\langle xDz|, \tag{2b}$$

$$\mathbf{P}_C = \sum_{xy} |xyR\rangle\langle xyR|, \quad \mathbf{P}'_C = \sum_{xy} |xyD\rangle\langle xyD|, \tag{2c}$$

where the subscripts indicate the agent performing the measurement and the summation is over any orthonormal basis.

The outcomes of each measurement are given by the corresponding Hermitian operators

$$\mathbf{X} = 2\mathbf{P}_A - \mathbf{1}, \quad \mathbf{X}' = 2\mathbf{P}'_A - \mathbf{1}, \tag{3a}$$
$$\mathbf{Y} = 2\mathbf{P}_B - \mathbf{1}, \quad \mathbf{Y}' = 2\mathbf{P}'_B - \mathbf{1}, \tag{3b}$$
$$\mathbf{Z} = 2\mathbf{P}_C - \mathbf{1}, \quad \mathbf{Z}' = 2\mathbf{P}'_C - \mathbf{1}, \tag{3c}$$

with eigenvalues $+1$ and $-1$.

If Alice, Bob, and Charlie all choose to measure in the *R/L* basis, then the expectation value of the product of their measurements is

$$\langle\psi|\mathbf{XYZ}|\psi\rangle = +1. \tag{4}$$

If, however, exactly one of them chooses to measure in the *D/A* basis, we instead have

$$\langle\psi|\mathbf{XY}'\mathbf{Z}'|\psi\rangle = \langle\psi|\mathbf{X}'\mathbf{YZ}'|\psi\rangle = \langle\psi|\mathbf{X}'\mathbf{Y}'\mathbf{Z}|\psi\rangle = -1 \tag{5}$$

For the other four possible measurement choices, the expectation value is zero, meaning that a product of $+1$ occurs just as often as a product of $-1$.

A difficulty seems to arise when one tries to interpret these predictions in terms of a local hidden-variable model. Suppose $a$ is the hidden variable for a given preparation of the state $|\psi\rangle$, and suppose there are functions $X$, $X'$ mapping the hidden-variable space $S$ to the set $\{-1,+1\}$ of possible outcomes such that $X(a)$ is the outcome Alice obtains if she measures in the *R/L* basis, $X'(a)$ is her outcome is she chooses the *D/A* basis. Suppose further that there are similar functions $Y$, $Y'$ for Bob and $Z$, $Z'$ for Charlie.

Since the expectation value in Eq. (4) attains its maximal value, it must be that almost every measurement of this sort results in a product of $+1$. Let $I_1$ denote the measurable subset of $S$ such that, for all $a \in I_1$, we have $X(a)Y(a)Z(a) = +1$. For consistency with the quantum predictions, we must have that a product of $+1$ occurs almost surely when all three agents measure in the *R/L* basis. In a similar manner, let $J_2, J_3, J_4$ be measurable sets such that $X(a)Y'(a)Z'(a) = -1$ for all $a \in J_2$, $X'(a)Y(a)Z'(a) = -1$ for all $a \in J_3$, and $X'(a)Y'(a)Z(a) = -1$ for all $a \in J_4$.

Now, let us be so bold as to assume that there exists at least one $a$ contained in all four sets $J_2, J_3, J_4$, and $I_1$. It follows from the fact that $a \in J_2 \cap J_3 \cap J_4$ that

$$X(a)Y(a)Z(a)\Big[X'(a)^2\,Y'(a)^2\,Z'(a)^2\Big] = -1, \tag{6}$$

**Table 1** Simple GHZ hidden-variable model

|       | X | X′ | Y | Y′ | Z | Z′ |
|-------|---|----|---|----|---|----|
| $a_1$ | − | 0  | − | 0  | + | 0  |
| $a_2$ | + | 0  | 0 | +  | 0 | −  |
| $a_3$ | 0 | −  | − | 0  | 0 | −  |
| $a_4$ | 0 | −  | 0 | −  | − | 0  |

Each row corresponds to one of four hidden-variable states $a_1,\ldots,a_4$. For each row, the six columns give the value of $X,\ldots,Z'$, where "+" is +1, "−" is −1, and "0" indicates a missed detection which, of course, contradicts the assumption that $a \in I_1$. We conclude, therefore, that consistency with the ideal quantum predictions implies that the intersection of the four sets is empty; i.e.,

$$I_1 \cap J_2 \cap J_3 \cap J_4 = \varnothing. \tag{7}$$

How can four sets, each corresponding to an event that occurs almost surely, be mutually exclusive? One way to understand this is to realize that the hidden-variable model must be contextual in the sense that the probability distribution over the hidden-variable states must vary depending upon which bases Alice, Bob, and Charlie choose to measure in [26–29]. Now, the locality condition would seem to rule out such contrived dependencies, but there is a simpler and more natural way in which contextuality may arise.

In reality, not all measurements result in an outcome; some result in missed detections. This simple fact allows for a local hidden-variable model to be contextual. Consider the simple local hidden-variable model shown in Table 1. There, we can see that $I_1 = \{\boldsymbol{a}_1\}$ and $J_i = \{\boldsymbol{a}_i\}$ for $i \in \{2, 3, 4\}$, so Eq. (7) is clearly satisfied. Whenever all three agents measure in the *R/L* basis and all three get a detection, a product of +1 is surely obtained. Similarly, when exactly one of them measures in the *D/L* basis (and all of them get a detection), a product of −1 is certain to result. Thus, by post-selecting on coincidence detections we are able to exhibit contextuality and achieve perfect agreement with the ideal quantum predictions.

The above example is clearly a mathematical contrivance, but it serves to illustrate the importance of missed-detection events. Normally, such events are seen as a consequence of the imperfections of physical devices, but this need not be so. The following section describes a somewhat more physically motivated local hidden-variable model that, while not reproducing exactly the ideal quantum predictions, does give results that appear more consistent with experimental observations.

## 3 Local hidden-variable model

Reference [24] describes a local hidden-variable model capable of violating the CHSH inequality under spacelike separated measurements when only coincidence detections are used. This model is loosely based on prior work in stochastic electrodynamics and, in particular, stochastic optics [30]. In this model, a quantum optics system is modeled as a combination of an experimenter-defined "signal" and an uncontrolled random "noise" term. Given a normalized quantum state $|\psi\rangle$ in an $N$-dimensional Hilbert space $\mathcal{H}$, we consider its complex components $\alpha_n = \langle n | \psi \rangle$ in the *H/V* basis. For the GHZ state, $\alpha_1 = 1/\sqrt{2}$, $\alpha_8 = -i/\sqrt{2}$, and all other components are zero. The aforementioned "signal" term is taken to be $s\boldsymbol{\alpha}$, a scaled version of the state vector, where $s \geq 0$. The noise term is given by $\boldsymbol{v} = \sigma \mathbf{z}/\|\mathbf{z}\|$, where $\sigma \geq 0$ and $\mathbf{z}$ is a zero-mean complex Gaussian random vector. The coherent sum of these two terms, $\boldsymbol{a} = s\boldsymbol{\alpha} + \boldsymbol{v}$, constitutes the complete hidden-variable state.

For $N = 2$ this model may be thought to represent a single photon as a classical electromagnetic plane wave, with Jones vector $s\boldsymbol{\alpha}$ for the electric field, that is coherently added to a single-wavevector mode of the vacuum zero-point field, with a random Jones vector $\sigma \boldsymbol{v}$. The parameter $\sigma$ defines the scale of the background field, notionally corresponding to $\sqrt{\hbar \omega}$, where $\hbar$ is Planck's constant divided by $2\pi$ and $\omega$ is the angular frequency. Thus, we may think of $s/\sigma < 1$ as corresponding to the quantum regime. We emphasize that this is a mathematical model, not a physical theory, but one that is perhaps not entirely without physical significance or insight.

Since we intend to describe a local, deterministic model, a procedure for determining measurement outcomes must be specified. This is done as follows: given a projection operator $\mathbf{P}$ defining a sharp measurement for some bivariate observable and a particular hidden-variable realization $\mathbf{a}$, we say that an outcome of 1 occurs if $\mathbf{a}^\dagger \mathbf{P} \mathbf{a} > \gamma^2$, for some fixed detection threshold $\gamma \geq 0$, and an outcome of 0 occurs if $\mathbf{a}^\dagger (\mathbf{1} - \mathbf{P}) \mathbf{a} > \gamma^2$; otherwise, a null result is said to occur. The motivation for this definition arises from the physics of photodetectors, which classically require a certain critical intensity over time in order to register a detection. For $s \leq (\sqrt{2} - 1)\sigma$ and $\gamma \geq \sigma$, it can be shown that a null result will only occur if $\mathbf{a}^\dagger \mathbf{P} \mathbf{a} \leq \gamma^2$ and $\mathbf{a}^\dagger (\mathbf{1} - \mathbf{P}) \mathbf{a} \leq \gamma^2$, so double detections are not possible [24]. Furthermore, no detections at all are possible for $\gamma \geq \sqrt{2}\sigma$. In what follows, we shall therefore take $s = (\sqrt{2}-1)\sigma$, $\gamma \in [\sigma, \sqrt{2}\sigma)$, and, for simplicity $\sigma = 1$.

The possibility of obtaining no outcome is in sharp contrast to the idealized notion of a quantum measurement, wherein an outcome of 1 or 0 is always obtained, with probability $\boldsymbol{\alpha}^\dagger \mathbf{P} \boldsymbol{\alpha}$ and $\boldsymbol{\alpha}^\dagger (\mathbf{1} - \mathbf{P}) \boldsymbol{\alpha}$, respectively. Nondetections are, however, quite commonplace in quantum optics. Though normally thought of as a result of attenuation and imperfect detectors, the present model treats nondetections as an intrinsic part of the underlying physical phenomena. As we have seen, this property is important for understanding the GHZ puzzle from a local hidden-variable perspective.

With this definition of measurement, we can now define random variables corresponding to the observables in the GHZ puzzle. For a given value of the threshold $\gamma$ and a give realization $\mathbf{a} \in S$, let $X(\mathbf{a})$ denote the outcome of Alice's measurement when performed in the $R/L$ basis, with nondetections simply reported as 0. Thus,

$$X(\mathbf{a}) = \begin{cases} +1 & \text{if } \mathbf{a}^\dagger \mathbf{P}_A \mathbf{a} > \gamma^2, \\ -1 & \text{if } \mathbf{a}^\dagger (\mathbf{1} - \mathbf{P}_A) \mathbf{a} > \gamma^2, \\ 0 & \text{otherwise.} \end{cases} \tag{8}$$

On the other hand, let $X'(\mathbf{a})$ denote the outcome Alice *would have obtained* if she had chosen the $D/A$ basis instead. In this case,

$$X'(\mathbf{a}) = \begin{cases} +1 & \text{if } \mathbf{a}^\dagger \mathbf{P}'_A \mathbf{a} > \gamma^2, \\ -1 & \text{if } \mathbf{a}^\dagger (\mathbf{1} - \mathbf{P}'_A) \mathbf{a} > \gamma^2, \\ 0 & \text{otherwise.} \end{cases} \tag{9}$$

We may similarly define $Y(\mathbf{a}), Y'(\mathbf{a})$ for Bob and $Z(\mathbf{a}), Z'(\mathbf{a})$ for Charlie. Within the context of the present local hidden-variable model, all these counterfactual quantities are well defined.

Let us now define the eight subsets $I_1, \ldots, I_8$ of $S$ corresponding to a product of $+1$ for each of the eight measurement choices $XYZ$, $XY'Z'$, $X'YZ'$, $X'Y'Z$, $X'YZ$, $XY'Z$, $XYZ'$, and $X'Y'Z'$, respectively. The subsets $J_1, \ldots, J_8$ are defined similarly but for products of $-1$. For a given measurement category $\mu \in \{1, \ldots, 8\}$, the union $I_\mu \cup J_\mu$ represents the set of all coincidence detections. Note that $I_\mu \cap J_\mu = \varnothing$, provided $\gamma \geq 1$.

The distribution of the hidden variable $\mathbf{a}$ defines a probability measure Pr, with which we can define the expected value $E_\mu$ of the product of the three measurements over all triple-coincidence detections for each measurement type $\mu$:

$$E_\mu = \frac{\Pr[I_\mu] - \Pr[J_\mu]}{\Pr[I_\mu] + \Pr[J_\mu]}. \tag{10}$$

The values of $E_\mu$ for a given $\mu$ and $\gamma$ can be estimated numerically. Using a random sample of $2^{20} \approx 10^6$ such realizations and a detection threshold of $\gamma = 1$, the following values were found: $E_1 = +0.86$, $E_2 = -0.87$, $E_3 = -0.85$, $E_4 = -0.86$ for the first four, and $E_5 = 0.006$, $E_6 = -0.01$, $E_7 = 0.0030$, $E_8 = 0.002$ for the last four, each with a standard deviation of about 0.01. The ideal quantum predictions are $E_1 = +1$, $E_2 = E_3 = E_4 = -1$ and $E_5 = \ldots E_8 = 0$. The first four of these correlations are plotted in Fig. 1 and compared against measured results from the experiments of Refs. [21–23]. In all cases the correlations predicted by the local hidden-variable model are seen to be comparable to or greater than those observed experimentally. Note that the correlations of Ref. [21] were actually of opposite sign.

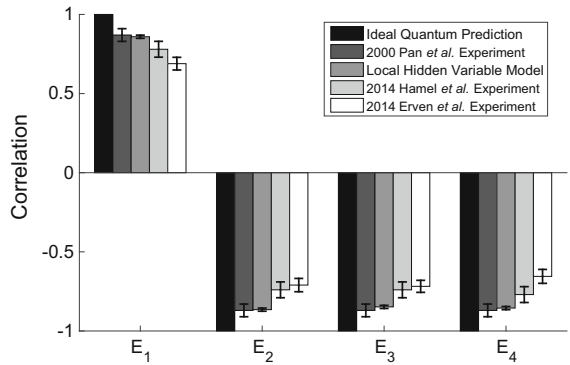

**Fig. 1** Plot of the correlation values $E_1, \ldots, E_4$ for the ideal quantum predictions (*black*), the 2000 Pan et al. experiment of Ref. [21] (*dark grey*), the local hidden-variable model with $\gamma = 1$ (*grey*), the 2014 Hamel et al. experiment of Ref. [23] (*light grey*), and the 2014 Erven et al. experiment of Ref. [22] (*white*). The *error bars* indicate statistical or measurement uncertainty

## 4 Threshold dependence

We have seen that the local hidden-variable model of the previous section exhibits strong correlations, even for a minimum detection threshold of $\gamma = 1$. In this section, we will consider the behavior of the model as the detection threshold is increased.

A quantity of particular interest in the Mermin statistic $M$, defined by

$$M = |E_1 - E_2 - E_3 - E_4|, \tag{11}$$

which takes on a maximal value of $M = 4$ under the ideal quantum predictions. In the 2014 Erven experiment of Ref. [22], a value of $M = 2.77 \pm 0.08$ was obtained, while the 2014 Hamel experiment of Ref. [23] obtained a value of $M = 3.04 \pm 0.10$. In the 2000 Pan experiment of Ref. [21], a value of $M = 3.48 \pm 0.08$ may be inferred. According to Mermin, any local hidden-variable model must yield a value of $M \leq 2$ [18], but sufficiently low detection efficiencies can permit violations of this inequality [19]. It is for this reason that the present local hidden-variable model is able to achieve a value of $M = 3.43 \pm 0.02$ that is well above the Mermin bound, albeit still below the ideal value of $M = 4$.

Interestingly, this agreement can be made better, even exact, by increasing the detection threshold $\gamma$. Figure 2 shows the Mermin statistic computed by Monte Carlo simulation using increasing values of $\gamma$. We find that $M$ increases monotonically with $\gamma$, reaching its maximal value around $\gamma = 1.09$. Thus, by placing more stringent requirements on detection, one can approach the ideal quantum predictions.

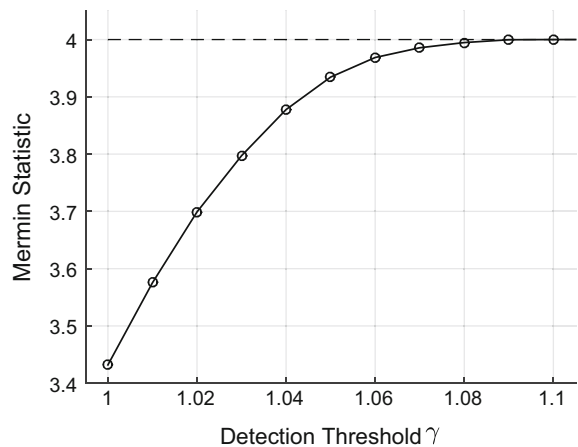

**Fig. 2** Plot of the Mermin statistic $M$ versus detection threshold $\gamma$ for the local hidden-variable model. The *dashed horizontal line* indicates the Mermin inequality upper bound. For large $\gamma$, the value of $M$ approaches 4

The price paid for this improved agreement is, of course, a loss of detection efficiency. Larsson gives four different definitions of detection efficiency for three-particle GHZ states [19]. The first, denoted $\eta_1$, is simply the number of single-photon detections over the total number of incident photons (more precisely, $\eta_1$ is defined as the minimum of this ratio over the six measurement-and-agent combinations; equivalently, over all eight measurement categories $\mu$). The second, $\eta_{2,1}$, is the probability of obtaining a second detection given a single "heralding" detection. For three-photon systems, the quantity $\eta_{3,2}$ is defined as the probability of obtaining a third detection given two other detections. Finally, $\eta_{3,1}$ is the probability of obtaining a second and third detection given a single detection. According to Theorem 2 of Ref. [19], any one of the following conditions will preclude a local hidden-variable model: $\eta_1 > 5/6$, $\eta_{2,1} > 4/5$, $\eta_{3,2} > 3/4$, or $\eta_{3,1} > 3/5$.

The definition of $\eta_1$ relies on a notion of discrete incident photons that, while conceptually appealing, is difficult to define operationally. In the present hidden-variable model, $\eta_1$ is both well defined and readily calculated. For example, the single-photon detection efficiency of Alice when measuring in the R/L and D/A basis is given by $\Pr[X \neq 0]$ and $\Pr[X' \neq 0]$, respectively, with similar expressions for Bob and Charlie. Thus, $\eta_1$ would be the minimum of these six values. Similarly, $\eta_{3,2}$ would be computed as the minimum over terms such as $\Pr[X \neq 0 | Y \neq 0, Z \neq 0]$, $\Pr[Y \neq 0 | X' \neq 0, Z \neq 0]$, etc., using all $3 \times 8 = 24$ combinations of agents and measurement categories. All of these quantities may be estimated by Monte Carlo sampling.

With $\gamma = 1$, it was found that only about 8.1 % of the realizations resulted in a detection for any given agent and measurement choice. Similarly, $\eta_{2,1}$ was found to be about 28 %, $\eta_{3,2}$ was about 36 %, and $\eta_{3,1}$ was about 10 %. Note that $\eta_{2,1} \neq \eta_1$, $\eta_{3,2} \neq \eta_1$, and $\eta_{3,1} \neq \eta_1 \eta_{2,1}$, so these results imply that the missed-detection events in this model are not independent, contrary to, say, the assumptions of Ref. [20].

All of these efficiency values are also well below the Larsson critical bounds, indicating that a local hidden-variable model would be expected to exist that is capable of violating the Mermin inequality. As we have seen, increasing the threshold tends to increase the Mermin statistic, but this also decreases the efficiencies. Interestingly, though, not all efficiencies are uniform across all agents and measurement choices. The six component efficiencies used to compute $\eta_1$ all appear similar, as do the 24 used to compute $\eta_{2,1}$, but those of $\eta_{3,2}$ and $\eta_{3,1}$ exhibit a curious bimodal clustering.

Figure 3 shows a plot of the triple coincidence efficiency $\eta_{3,2}$ as a function of detection threshold $\gamma$. As expected the efficiency decreases with increasing threshold, but there is a distinct separation among the 24 combinations of three agents and eight measurement choices used to determine $\eta_{3,2}$. Combinations of measurement choices used in the GHZ puzzle, such as $X, Y, Z$ and $X, Y', Z'$, exhibit a distinctly larger detection efficiency over the others yet are quite similar among themselves. Defining $\eta'_{3,2}$ to be the minimum over these 12 combinations, we find that $\eta'_{3,2} > \eta_{3,2}$, with increasing relative separation as $\gamma$ is increased. The example of Table 1 shows an extreme example of this, with $\eta'_{3,2} = 1$ and $\eta_{3,2} = 0$. It would be curious, then, to see if such behavior is exhibited in the real world.

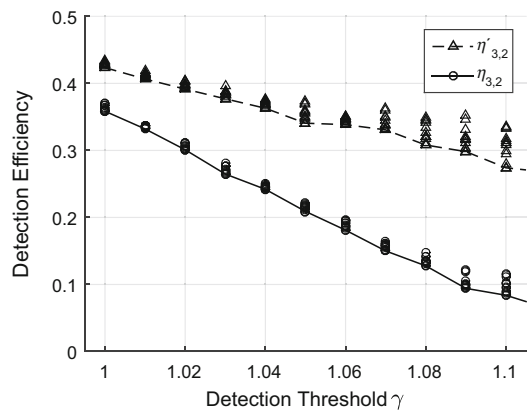

**Fig. 3** Plot of the efficiencies $\eta_{3,2}$ and $\eta'_{3,2}$. The *solid line* shows the minimum over all 24 combination of agents and measurement choices. The *dashed line* shows the minimum over just those combinations corresponding to the four measurement selections appearing in the GHZ puzzle. The *circles* and *triangles* correspond to the detection efficiencies for each of the 24 combinations, with *triangles* denoting those used to compute $\eta'_{3,2}$ and *circles* denoting the remaining 12 combinations

## 5 Conclusions

A local hidden-variable model of the GHZ puzzle has been presented and shown to give predictions in accordance with experimental observations. A key feature of this model is that many measurements result in a null outcome. The resulting missed detections give efficiencies that are within the bounds required for consistency with quantum predictions. The hidden variable is modeled as the sum of a fixed "signal" term specified by the prepared state and random "noise" term representing quantum fluctuations. The amplitude of the coherent sum of these two terms, when compared against a fixed threshold, determines the outcome of a measurement. The three measuring agents each have the same coherent sum but are free to choose any basis in which to measure, so the model is locally causal. Each realization, then, may be more favorable to detection for certain measurement choices than for others. Thus, within this model, missed detections are not the result of independent errors associated with the detector but are a determininstic consequence of the coherence properties of the hidden-variable state. This provides an interesting alternative viewpoint to understanding photon detection efficiency.

Raising the detection threshold places a more stringent requirement for detection, and this, too, entails some interesting observations. First, we find that the model predicts better agreement with the ideal quantum predictions as the threshold is increased. Indeed, for sufficiently high thresholds, perfect agreement may be achieved. This comes, of course, at the price of lower efficiency. One finds an analogous situation in the case of photons transmitted over long distances, which, due to attenuation, exhibit significant losses. Those losses are commonly viewed as discrete events (i.e., a photon is either transmitted or lost); here we offer an alternative interpretation that the attenuation is real and continuous, thereby resulting in fewer threshold exceedances.

Finally, we note an interesting observation that may have real-world consequences. According to this model, detection events are not independent and may differ across measuring agents and basis choices. In particular, we find that for GHZ states the measurement choices corresponding to those bases giving strong correlations tend to have a higher triple-detection efficiency that those expected to give little or no correlation. If this turns out to be true experimentally, it will have important implications for understanding of quantum measurement.

**Acknowledgments** This work was supported by the Office of Naval Research under Grant No. N00014-14-1-0323. The author would like to thank J. Troupe, Q. Niu, and K. Resch for their helpful discussions and comments on the manuscript.